\documentclass[iop]{emulateapj} 

\usepackage{apjfonts}
\usepackage{bm}
\usepackage{CJK}
\usepackage{soul}
\usepackage{color}
\usepackage{soul}

\newcommand{\set}[1]{\bm{#1}}
\newcommand{\starlabel}{\ell}
\newcommand{\labels}{\set{\starlabel}}

\newcommand{\Hessian}{\underline{\underline{\mathbf{H}}}}
\newcommand{\ai}{{\it a.i.~}}

\shorttitle{Constructing Polynomial Spectral Models for Stars}
\shortauthors{Rix, Ting, Conroy, Hogg}

\begin{document}

\begin{CJK*}{UTF8}{gbsn}
\title{Constructing Polynomial Spectral Models for Stars}

\author{Hans-Walter Rix\altaffilmark{1}, Yuan-Sen Ting (丁源森)\altaffilmark{2}, Charlie Conroy\altaffilmark{2}, David W. Hogg\altaffilmark{1,3,4,5}}

\altaffiltext{1}{Max Planck Institute for Astronomy, K\"onigstuhl 17, D-69117 Heidelberg, Germany}
\altaffiltext{2}{Harvard--Smithsonian Center for Astrophysics, 60 Garden Street, Cambridge, MA 02138, USA}
\altaffiltext{3}{Simons Center for Data Analysis, 160 Fifth Avenue, 7th floor, New York, NY 10010, USA}
\altaffiltext{4}{Center for Cosmology and Particle Physics, Department of Physics, New York University, 4 Washington Pl., room 424, New York, NY 10003, USA}
\altaffiltext{5}{Center for Data Science, New York University, 726 Broadway, 7th floor, New York, NY 10003, USA}

%
%
%
%
%
%
\begin{abstract}
Stellar spectra depend on the stellar parameters and on dozens of photospheric elemental abundances. Simultaneous fitting of these $\mathcal{N}\sim 10-40$ model labels to observed spectra has been deemed unfeasible, because the number of {\it ab initio} spectral model grid calculations scales exponentially with $\mathcal{N}$. We suggest instead the construction of a polynomial spectral model (PSM) of order $\mathcal{O}$ for the model flux at each wavelength. Building this approximation requires a minimum of only ${\mathcal{N}+\mathcal{O}\choose\mathcal{O}}$ calculations: e.g. a quadratic spectral model ($\mathcal{O}=2$) to fit $\mathcal{N}=20$ labels simultaneously, can be constructed from as few as 231 {\it ab initio} spectral model calculations; in practice, a somewhat larger number ($\sim 300-1000$) of randomly chosen models lead to a better performing PSM. Such a PSM can be a good approximation only over a portion of label space, which will vary case by case. Yet, taking the APOGEE survey as an example, a single quadratic PSM provides a remarkably good approximation to the exact {\it ab initio} spectral models across much of this survey: for random labels within that survey the PSM approximates the flux to within $10^{-3}$, and recovers the abundances to within $\sim 0.02$ dex {\it rms} of the exact models. This enormous speed-up enables the simultaneous many-label fitting of spectra with computationally expensive {\it ab initio} models for stellar spectra, such as non-LTE models. A PSM also enables the simultaneous fitting of observational parameters, such as the spectrum's continuum or line-spread function.
\end{abstract}


%
%
%
%
%
%

\section{Fitting stellar spectra}
\label{sec:introduction}

The spectra of stars encode an enormous amount of information, mainly about the stars' current physical state and the composition of the chemical elements in their photosphere. But the number of stellar labels\footnote{We use the term ``labels'' to mean the union of stellar parameters and photospheric elemental abundances, because in the current context these two classes of stellar attributes are being treated equivalently.} that fully specify a spectrum is large: a handful of stellar parameters and much of the periodic table. We know that stellar spectra with S/N$\,\sim 100$ and $R\sim 20$,$000-40$,$000$, currently emerging for $10^{4-6}$ objects from various surveys, contain the information to constrain 10--40 labels, at least for stars with favorable effective temperatures, $\sim 4000\,$K--$7000\,$K \citep[e.g.,][]{smi14,gar15,she15}. The accuracy and precision of label estimates for vast stellar samples matters greatly for understanding the formation of the Galaxy, stellar physics, and the origin of the chemical elements \citep[e.g.,][] {fre02,rix13,fre15}. 

A principled determination of these stellar labels requires to fit the data with physical model spectra, in which the stellar labels constitute 10--40 model parameters. The calculation of such {\it ab initio} spectral models (\ai models) through radiative transfer calculations has a storied tradition \citep[for an overview, see][]{smi14,gar15}. Current \ai models vary by the degree of physical simplification they apply: LTE {\it vs} non-LTE; plane-parallel {\it vs} spherical geometry; 1D, averaged or full 3D; static {\it vs} time dependent; and by the extent and robustness of the atomic data that underlie them.

The computation of \ai models is expensive, all the more so if the simplifying assumptions are dropped. This is why ``brute force'' fitting of spectra with \ai models (of, say, 10--40 labels) is unfeasible for the foreseeable future: most approaches to fitting \ai models to observed spectra have relied on pre-computing grids of \ai spectra in the $\mathcal{N}$-dimensional label space, and then interpolating between them pixel-by-pixel, e.g., quadratically (i.e., $2^{\rm nd}$-order) or cubic (i.e., $3^{\rm rd}$-order) as in \citet{all06,all14}. But for any number of grid points, $M\approx 3-5$, in each label-dimension, the total number of \ai model calculations required grows exponentially with the dimension $\mathcal{N}$ of label space: $N_{\rm tot}\propto M^\mathcal{N}\propto \exp{(\mathcal{N}\cdot\ln{M})}$. Established approaches have coped with this in practice by fitting models spectra first in a 3--6 dimensional sub-space of $\mathcal{N}$, and subsequently fitting one (or two) further label at a time, holding the initial labels fixed. This approach has important limitations with with state of-the-art data: first, \citet{tin16} (hereafter T16) has shown that more than just 2 or 3 elemental abundances affect the atmosphere structure, and hence are physically covariant with the basic stellar parameters; second, physical correlations and data-driven covariances are known to exist among (abundance) labels, but cannot be estimated when fitting one label at a time; third, to mitigate against unaccounted covariances, established fitting approaches have often focused on unblended lines, thereby under-exploiting the information content of the data by a large factor (T16).

T16 proposed a way to overcome this impasse by employing more linear algebra in the fitting, to save on \ai model calculations; in this {\it Letter} we take this idea a step further. T16 proposed to tessellate the space of stellar labels into a finite set of regions (dubbed linear Taylor-spheres, or 1OTS). Within each 1OTS the \ai model flux at each wavelength can be described sufficiently well by a linearized spectral model (LSM), linearized (in all labels) around the \ai model spectrum at a fiducial label value  \citep[see also][] {rec06}. T16 showed that such LSM can sufficiently approximate the exact model spectra within a 1OTS. Together with the finite number of Taylor-spheres, required to cover any given spectral survey (e.g., $\sim 150$ for the APOGEE red clumps), this leads to a dramatic reduction in the total number of \ai model calculations: simultaneous fitting of 10--40 labels should then be feasible.

Here we point out a rather obvious extension of this idea, which yields even greater computational savings: the construction of approximate model spectra, where the predicted flux at each pixel by a polynomial in all labels away from a fiducial model spectrum. This idea had been put forth by \citet{pru11} for empirical spectra, who did, however, not pursue its potential of fitting many labels simultaneously. We denote such approximate {\it polynomials spectral models} as PSM, to distinguish them from the \ai models themselves. It is important not to think of these PSM as a $\mathcal{O}^{th}$-order interpolation between a {\it pre-calculated} grid of \ai models (as e.g., \citet{pru11} did for a quadratic PSM in three labels), as this would still require $M_{grid}^\mathcal{N}\propto\exp{(\mathcal{N}\cdot\ln{M_{grid}})}$ \ai model calculations. Instead, one should think of determining the (near)-smallest number of \ai model spectra (specified by $\mathcal{N}$ labels) one needs to calculate in order to construct a $\mathcal{O}^{th}$-order approximation to the \ai model spectra. The simplification and speed-up of such spectral fitting compared to T16 arise from the fact that a single PSM can approximate the \ai model spectra over a much larger volume in label space. While this shares the idea of a polynomial flux approximation with {\it The Cannon} \citep{nes15a}, it is {\it not} data-driven model building.

In the subsequent Sections we first derive that the minimal number of \ai models needed to construct a PSM of order $\mathcal{O}$ and then illustrate heuristically how well, and over what volumes in label space, these PSMs approximate the \ai models.

%
%
%
%
%
%

\section{A polynomial model approximation for ab initio model spectra of stars}

Following T16, we suppose that an \ai modeling ``machinery'' can predict the normalized flux of a synthetic spectrum, $\mathbf{f}_{a.i.} (\lambda|\labels )$, given a set of stellar labels, $\labels$. We {\it assume} that the \ai model spectra change from point to point in label-space, but do so smoothly or differentiably at every wavelength. Then the \ai model spectrum at any $\labels$ sufficiently close to an model grid point $\labels_*$ (within a $1^{\rm st}$-order Taylor-sphere or 1OTS, in the nomenclature of T16) can therefore be described with high accuracy by a {\it linear} spectral model (LSM, see T16):

\begin{equation}
\mathbf{f}_{\rm lin}(\lambda|\labels_* + \mathbf{\Delta}\labels) \simeq \mathbf{f}_{a.i.}(\lambda|\labels_*) + \mathbf{\Delta}\labels^T \cdot \overrightarrow{\mathbf{g}} (\lambda|\labels_*) ,
\label{eq:linear}
\end{equation}

\noindent 
where $\overrightarrow{\mathbf{g}} (\lambda|\labels_*) \equiv \overrightarrow{\nabla}_{\labels} \mathbf{f}_{a.i.} (\lambda|\labels_*)$.

In principle, specifying a LSM merely requires $1+\mathcal{N}$ model calculations, but T16 showed a a factor of a few more is needed to explore the actual extent of the 1OTS. This LSM approximation, $\mathbf{f}_{\rm lin}$, can obviously be generalized to a polynomial spectra model (PSM):

\begin{eqnarray}
\mathbf{f}_{\rm PSM}(\lambda|\labels_* + \mathbf{\Delta}\labels) \simeq \mathbf{f}_{a.i.}(\lambda|\labels_*) + \mathbf{\Delta}\labels^T \cdot \overrightarrow{\mathbf{g}} ~(\lambda|\labels_*) \nonumber \\
~+~ \bold{\Delta}\labels^T \cdot \Hessian ~(\lambda|\labels_*) \cdot \bold{\Delta}\labels~+~...~,
\label{eq:quadratic}
\end{eqnarray}

\noindent
where we will focus on $2^{\rm nd}$-order, both for astrophysical reasons (it may work well enough) and to avoid cumbersome notation. Such a PSM holds for every one of the $K$ wavelengths $\lambda$. One may think of it as a model with $K$ $0^{th}$-order terms, $\mathbf{f}_{\rm PSM}(\lambda_k|\labels_*)$, then $K\times \mathcal{N}$ $1^{\rm st}$-order terms, and finally $K\times \mathcal{N} (\mathcal{N} + 1)/2$ ~$2^{\rm nd}$-order terms. The number $\mathcal{N}(\mathcal{N} + 1)/2$ arises because of the symmetry of $\Hessian$. In total that makes for

\begin{equation}
K\times N_{\rm tot}\equiv K\times \bigl ( 1+\mathcal{N}
~+~\mathcal{N}\cdot (\mathcal{N}+1)/2\bigr )
\end{equation}

\noindent
unknown terms. For more general PSM of order $\mathcal{O}$, one has $K\times N_{\rm tot}=K\times{\mathcal{N}+\mathcal{O}\choose\mathcal{O}}$.

If we compute \ai models $\mathbf{f}_{a.i.}(\lambda|\labels_* + \mathbf{\Delta}\labels)$ at ${\mathcal{N}+\mathcal{O}\choose\mathcal{O}}$ different points in label space, $\mathbf{\Delta}\labels$, we have created exactly $K\times N_{\rm tot}$ left-hand-side terms to solve exactly for the terms that specify the PSM. Note that strictly speaking $\overrightarrow{\mathbf{g}} ~(\lambda|\labels_*)$ and $\Hessian ~(\lambda|\labels_*)$ are not exactly the ``gradient'' and the ``Hessian'', but merely the $1^{\rm st}$ and $2^{\rm nd}$-order coefficients that solve the equation.

Compared to the 1OTS, we have to calculate $1+\mathcal{N}/2$ times more \ai models for any one quadratic PSM. But if the the region in label space around $\labels_*$ over with this quadratic PSM works is sufficiently larger, an important speed-up over the the (set of) LSM should result. Calculating somewhat more \ai models than this minimum, and solving Eq.\ref{eq:quadratic} in a least squares sense, makes for a much better conditioned solution for a PSM, as we show below.

\begin{figure*}
\centering
\includegraphics[width=0.78\textwidth]{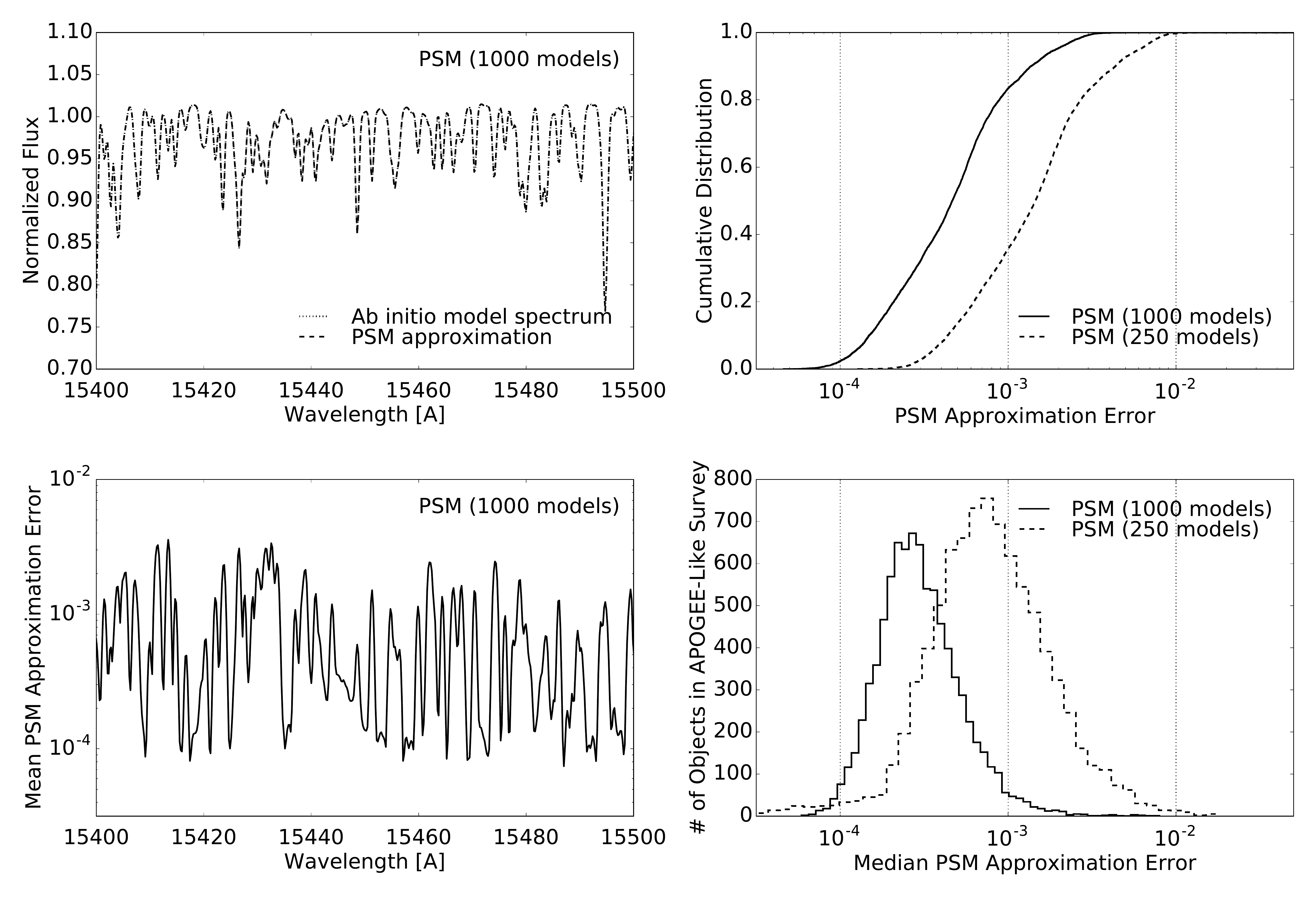}
\caption{
Quality of the (quadratic) PSM approximation: a single PSM was constructed using 250 or $1$,$000$ {\it a.i.} model spectra ({\it cf.} the absolute minimum number of 231), calculated at label points (``objects'') drawn randomly from those in the APOGEE survey \citep{ala15,hol15}. The panels illustrate different PSM -- {\it a.i.} model comparisons, for $10$,$000$ other objects drawn from the labels of the APOGEE survey. The top left panel shows for a limited wavelength section the average of the exact {\it a.i.} model spectra and of the PSM, which appear indistinguishable. The bottom left panel shows the ensemble average (absolute) difference between the {\it a.i.} model and the PSM flux (the approximation error), as a function of wavelength. For each one of the $10$,$000$ objects there is a pixel-by-pixel distribution of these approximation errors, which is shown in the top right panel for the pixel-by-pixel average approximation error. The bottom right panel finally shows the distribution across all objects of their (pixel-by-pixel) median approximation error. Note that there are rare cases (objects of very high [Fe/H], where the approximation is only good to a median of $10^{-3}$. Taken together, however, this shows that a single PSM approximates the exact {\it a.i.} model spectra typically to within $10^{-3}$ for objects with a label distribution resembling that of the entire APOGEE survey (which merely serves as an illustration here), over the $10$,$000$ labels of the median. Constructing the PSM from 1000 instead of 250 random label points leads to a better PSM approximation.
}
\label{fig:rms}
\end{figure*}
\vspace{0.5cm}

%
%
%
%
%
%

\section{Verification of PSM accuracy, using Kurucz models for APOGEE-like spectra}

Strictly verifying the validity of the PSM approximation, like any approximation to a high-dimensional function, would be of enormous computational expense. Here, too, escaping the curse of dimensionality comes at a price: relying on the physically plausible {\it assumption} that spectral flux changes can be approximated by polynomials for modest label changes; and settling for heuristic and approximate ways to explore the extent in label space over which a single PSM is useful.

As in T16, we can set out for a pragmatic test of the PSM approximation, using Kurucz model spectra that resemble in resolution and wavelength coverage the APOGEE spectra; the arguments should hold qualitatively for other surveys, but need to be tested case-by-case. In total, the DR12 data release of the APOGEE \citep{ala15,hol15} provides 17 labels for each star ($T_{\rm eff}$, $\log g$ and 15 elemental abundances) while fixing $v_{\rm macro} = 6\,$km/s and adopting a $\log g-v_{\rm turb}$ relation for $v_{\rm turb}$. A quadratic PSM for 19 labels requires a minimum $N_{\rm tot}=210$ \ai model calculations. We chose the reference label, $\labels_*$, to be the APOGEE DR12 sample median in each of the 19 labels, providing $\mathbf{f}_{a.i.}(\lambda|\labels_*)$ in Eq.\ref{eq:quadratic}. The vast majority of targets in APOGEE are disk stars with all [X/H]$\,>-1$, and we restrict our PSM verification to this regime. We then drew 209 $\bold{\Delta}\labels$ at random from the APOGEE DR12 catalog. For the labels $v_{\rm turb}$ and $v_{\rm macro}$ we adopted the same $\log g- v_{\rm turb}$ relation from APOGEE with a spread of $0.2\,$km/s, and a distribution in $v_{\rm macro}$ uniform across $3\,$km/s -- $8\,$km/s. We convolved spectra to the APOGEE resolution assuming the combined LSF from APOGEE and using codes from the {\sc apogee}  Python package \citep{bov16}, and continuum normalized spectra the same way as {\it The Cannon} \citep{nes15a}. This provided the remaining 209 left-hand sides of Eq.\ref{eq:quadratic} to solve exactly for the $\overrightarrow{\mathbf{g}} ~(\lambda|\labels_*)$ and $ \Hessian ~(\lambda|\labels_*) $, fully specifies $\mathbf{f}_{\rm PSM} (\lambda|\labels_* + \mathbf{\Delta}\labels) $ from Eq \ref{eq:quadratic}.

As expected by construction of the PSM, $\mathbf{f}_{\rm PSM}(\lambda|\labels_* + \mathbf{\Delta}\labels) $ matches the {\it a.i.} model at all the 210 $\bold{\Delta}\labels$ exactly. This minimally constructed PSM also provides good approximations to $\mathbf{f}_{a.i.} (\lambda|\labels_* + \mathbf{\Delta}\labels)$ for other $\mathbf{\Delta}\labels$. Empirical experimentation showed that slightly over constraining Eq \ref{eq:quadratic} worked better: we calculated $\mathbf{f}_{a.i.}(\lambda|\labels_* + \mathbf{\Delta}\labels) $ for 250 and 1000 $\mathbf{\Delta}\labels$ drawn from APOGEE, and solved for the right hand side of Eq.\ref{eq:quadratic} in a least squares sense to determine the PSM coefficients.

There are two ways in which one can quantify how well the PSM, $\mathbf{f}_{\rm PSM}(\lambda|\labels_* + \mathbf{\Delta}\labels)$, approximates $\mathbf{f}_{a.i.}(\lambda|\labels_* + \mathbf{\Delta}\labels)$ for any $\mathbf{\Delta}\labels$ drawn from APOGEE: how well do the fluxes match, e.g.in a mean absolute deviation? And, at what accuracy level does the PSM approximation affect the label recovery? 

\begin{figure*}
\centering
\includegraphics[width=0.76\textwidth]{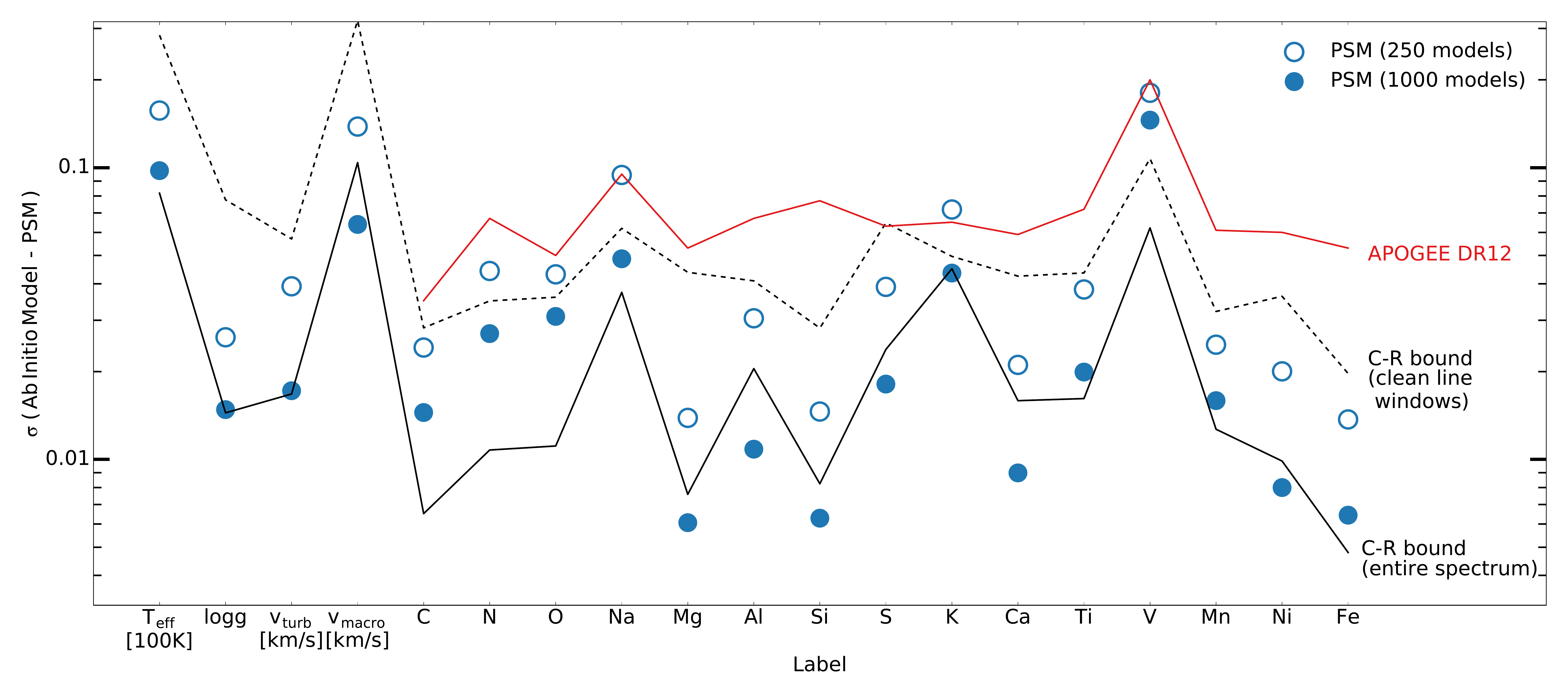}
\caption{Quality of the label recovery using the same PSM approximation as in Fig.\ref{fig:rms}, based on 250 (open circles) and 1000 (full circles) \ai model calculations, respectively. Shown is the {\it rms} difference between the labels of the PSM approximation that best matches the exact {\it a.i.} model spectrum in a $\chi^2$-sense, and the actual labels of the exact spectrum: PSM- induced errors in the label recovery by the PSM approximation are typically 0.02~dex (when considering the label range of the APOGEE survey). The dashed and solid lines show the theoretically achievable label precision at S/N$\,=100$ (the Cramer-Rao bound; see T16), when using the APOGEE wavelength windows, or the full spectrum. A single PSM approximation can be used for fitting all labels simultaneously across much of the APOGEE survey, without inducing serious systematic errors. The red line indicates typical APOGEE DR12 precisions. The quality of the label recovery remains (to within $\sim 10\%$ of each label's accuracy), even if a number of spectral continuum and line-spread parameters are also fit simultaneously.}
\label{fig:label}
\end{figure*}

Fig.\ref{fig:rms} illustrates how well an {\it a.i.} model spectrum of random star in APOGEE is can be approximated by the PSM in a mean absolute deviation sense. On average the PSM-predicted flux at any wavelength for a random star within APOGEE is within $10^{-3}$ or $10^{-3.5}$ of that for its {\it a.i.} model spectrum, depending on whether we used 250 or 1000 {\it a.i.} model calculations to construct the PSM. Fig \ref{fig:label} shows how much (or, how little) the PSM approximation, calculated here on the basis of 250 or 1000 \ai models, affects the label recovery across an APOGEE-like survey. The labels were recovered by a least squares fit of the PSM to noiseless \ai models, fitting all 19 labels {\it simultaneously}. These were then compared to the actual labels of the respective \ai models. With a single PSM, most labels are recovered as accurate as claimed precisions of current spectral surveys. More quantitatively the quality of the PSM label recovery is well-tracked by the information content that the spectra contain about any one label: following T16, this is quantified by the Cramer-Rao bound (for S/N$\sim 100$) using either only the APOGEE wavelength windows for certain elements or the whole spectrum.

The PSM appears heuristically as a better approximation when calculated on the basis of more {\it a.i.} model calculations, presumably for two reasons: the system of linear equations in Eq.\ref{eq:quadratic} becomes better conditioned; and a better sampling of label-space better mitigates any break-down of the polynomial approximation. Both factors must play a role: when we restrict the label range over which we first construct and then test the PSM, the PSM label recovery is even closer to the exact solution. Yet, the PSM constructed on the basis of 1000 (compared to 250) {\it a.i.} model calculations is still performing better. How many models to calculate for the PSM construction, and over which portion of label space to apply it will therefore depend in practice on the computational expense of the {\it a.i.} models and the desired label accuracy. Nonetheless, Fig.\ref{fig:rms} and Fig.\ref{fig:label} demonstrate that with calculating only 250 (or 1000) {\it a.i.} models one can construct a {\it single} (quadratic) PSM that performs remarkably well in approximating results from exact model spectra at random 19-dimensional label location across much of APOGEE survey.

%
%
%
%
%
%

\section{Prospects and limitations}

We have shown the advantages for spectral model fitting of generalizing the local linear expansion of {\it a.i.} model spectra laid out in T16 to higher order, constructing polynomial spectral models (PSM) that approximate the variations of the predicted spectral flux at each wavelength as a polynomial function of the labels. This reduces the calculation of the model spectra needed in simultaneous fitting of {\it many} stellar labels to observed spectra to linear algebra. Compared to established approaches that first calculate grids and then interpolate, the dramatic gain in constructing a PSM comes from the much more benign scaling of the computational effort with increasing label dimension: $\propto {\mathcal{N}+\mathcal{O}\choose\mathcal{O}}$, or $\propto \mathcal{N}^2$ for a quadratic model with $\mathcal{O}=2$, as opposed to $\propto \exp{(\mathcal{N}\cdot \ln M)}$. The way these PSM are constructed are mathematically very much analogous to data-driven {\it The Cannon} \citep{nes15a}, where a quadratic spectral model is derived form observed spectra. The arguments here provide a systematic guidance for the size of the required training set in {\it The Cannon}: we should expect the training set size to scale as (as a multiple) ${\mathcal{N}+2\choose 2}$, or $\propto \mathcal{N}^2$; this makes it plausible that {\it The Cannon} could constrain 19 labels from a training set of $10$,$000$ \citep{cas16}.

The heuristic verification of the PSM approximation, along with the framework laid out in T16, means that there should be no longer serious technical obstacles to determining stellar labels in large surveys to what amounts to fitting all labels with {\it a.i.} model spectra simultaneously. The accommodation of label correlation facilitates the extraction of abundance information from blended spectral features. We find from the gradient spectra, that 80\% of the spectrum's information on a label is spread over typically 30\% of all pixels, and is not just in narrow spectral windows (T16). For any given data set this should allow higher precision and accuracy. PSM also allows to treat parameters of the experimental set-up, such as the continuum fit or the spectral line-spread function (LSF) quasi as stellar labels, and fit them simultaneously. 

Of course, constructing PSMs is not a {\it panacea}: while a single PSM appears to suffice for the much of APOGEE survey, this is presumably because APOGEE has targeted stars in a rather restricted portion of label space: giant stars in a narrow temperature range. Yet, even there, constructing a separate PSM for the metal-poor regime may be advisable, as small model flux differences cause larger label recovery errors. Second, it is probably worth exploring the PSM approach to higher order in at least some of the labels. Perhaps most importantly, any fitting based on \ai spectral models can only work as well as the physics behind them. Insufficient atomic data or the restrictions of the LTE approximation remain untouched by the ideas laid out here. Nonetheless, we feel that T16 and this paper lay out a path that may help in doing justice to the enormous information content of present and future stellar spectroscopy surveys.

%
%
%
%
%
%

\acknowledgments
YST acknowledges support from NASA grant NNX15AR83H and is grateful to the MPIA and the DFG through the SFB 881 (A3) for their hospitality and financial support. CC acknowledges support from NASA grant NNX13AI46G, NSF grant AST-1313280, and the Packard Foundation. HWR's research contribution is supported by the ERC Grant Agreement n.$\,$[321035]. The computations in this paper were partially run on the Odyssey cluster supported by the FAS Division of Science, Research Computing Group at Harvard University. The computational work also used the Extreme Science and Engineering Discovery Environment (XSEDE), which is supported by National Science Foundation grant number ACI-1053575. We thank Bob Kurucz for developing and maintaining programs and databases without which this work would not be possible.

%
%
%
%
%
%
\end{CJK*}

\end{document}